\documentstyle[preprint,aps,12pt]{revtex}

\tighten

\begin{document}

\draft


\title{$U_A(1)$ symmetry restoration in QCD with $N_f$ flavors}

\author{Su H. Lee$^*$ and T. Hatsuda }

\address{Institute of Physics, 
 University of Tsukuba, Tsukuba, Ibaraki 305, Japan}

\address{$^*$Department of Physics, Yonsei University, Seoul 120-749, Korea}

\maketitle

\begin{abstract}

Recently, there have been reports that the chirally restored phase of 
QCD is effectively symmetric under $U(N_f) \times U(N_f)$ rather than
 $SU(N_f)\times SU(N_f)$.  We supplement their argument
by including the contributions from topologically nontrivial gauge field
configurations and discuss how the conclusions are modified.  
General statements are made concerning the particle spectrum of QCD
with light $N_f$ flavors in the high temperature chirally restored phase.

\end{abstract}

\vspace{1cm}


In relation to the exciting possibilities of probing the QCD phase transition
in relativistic heavy ion collisions, there have been an increasing
 number of studies on the nature and physical consequence 
of chiral symmetry restoration in QCD\cite{Gen}.
 Of particular interests are the possible 
 changes in meson properties such as masses, decay width or the 
spectral density in general as they might be observed directly through 
particle multiplicity or dilepton spectrum \cite{lep}.

Recently, there have been reports that the chirally restored phase of 
QCD is effectively symmetric under $U(N_f) \times U(N_f)$ rather than
 $SU(N_f)\times SU(N_f)$\cite{Shuryak1,Tom1}.
In particular, Cohen \cite{Tom1} has shown, from a general consideration 
of the QCD partition function, that the two point function of the 
 spin zero  multiplet of $U(N_f) \times U(N_f)$, namely 
 $\pi,\sigma,\delta,\eta'$ etc.,   
 become degenerate if the spontaneously broken chiral symmetry is 
restored at high temperature.  However, it should be noted that 
the argument in ref.\cite{Tom1} 
is not complete in the sense 
 that it does not take into account contributions from
topologically non-trivial gauge field  configurations.
In this letter, we supplement the work in \cite{Tom1} by including the
contributions from the non-trivial topological sectors and 
give general statements on the nature of chiral symmetry restoration of
QCD with light $N_f$ flavors.
       
Let us start with the Euclidean partition function of QCD. 

\begin{eqnarray}
Z[J]=\int D[A] D[\Psi \bar{\Psi}]
 {\rm exp}[-S_{QCD}-\bar{\Psi} J \Psi] \ ,
\end{eqnarray}
where $S_{QCD}=\frac{1}{4}F^2+\bar{\Psi} \rlap{/}D \Psi$.  
 $J$ could represent an external source or the mass matrix.
Here we will neglect the gauge fixing and ghost terms which will not
be relevant for our discussion.
 Consider integrating out the quark fields,

\begin{eqnarray}
\label{par2}
Z[J]=  \int D[A] 
e^{-S_{YM}}{\rm Det}[\rlap{/}{D} + J],
\end{eqnarray}
where $S_{YM} \equiv {1 \over 4}F^2$.
The integration over gauge fields $A$ can have topologically non-trivial
configurations with the topological charge $\nu=\frac{g^2}{32 \pi^2}
\int d^4x  F \tilde{F} \neq 0$:
\begin{eqnarray}
Z[J]=\sum_{\nu} Z[J]_{\nu}\ \ .
\end{eqnarray}
 
The configuration with non-zero $\nu$ is always accompanied by
 $n_+ (n_-)$ 
number of right-handed (left-handed) fermion zero mode satisfying
  $\rlap{/}D \psi_0=0$.  Furthermore, 
 the index theorem tells that $\nu = n_+ - n_-$.
  Our arguments below 
 are not limited to specific gauge configuration such as
 the BPST instanton  \cite{Bel1}.
 Topological
 gauge configurations  ($\nu \neq 0$) 
  have zero measure in the chiral limit ($J=0$)
as can be seen from the fermion determinant in eq.(\ref{par2}).
However, in the presence of external source or finite quark mass, these
zero-mode configurations can have non-trivial contribution.

 In the presence of 
one 
topological charge, the zero mode contribution can be separated out from
the fermion determinant and 
the $\nu=1$ sector of the  partition function becomes \cite{note}

\begin{eqnarray}
\label{par3}
Z[J]_{\nu=1}=  \int D[A]_{\nu =1} 
e^{-S_{YM}}{\rm Det}'[\rlap{/}{D}+ J]  \times  
  det_{st}(\int d^4x \bar{\psi}_0(x) J_{st}\psi_0(x)).
\end{eqnarray}

Here, $det_{sf}$ stands for the determinant in the flavor space which is
denoted by the index $s,f=1, \cdot \cdot 
 N_f$.  The rest of the fermion determinant
is denoted by ${\rm Det}'$,  which now does not have a zero mode contribution.
 $\psi_0 $ corresponds to
the zero mode solution in the presence of a nontrivial 
 gauge field configuration of $\nu=1$. 
 The explicit form of the functional integral
 for the case of the BPST instanton solution  has 
been worked out by  'tHooft \cite{tHooft1}.  Similar study for
 the periodic instanton relevant to finite temperature
 system is given in \cite{GPY}.

Given the tools, we will now study n-point functions at arbitrary temperature,
paying special attention to the contributions from the non-trivial  
topological  sector and its relation to
chiral symmetry restoration.

 For simplicity,  let us choose $N_f=2$. The case with more
flavors can be easily generalized, although they have important differences
as we will show later.
We will also restrict our formulas to include only $\nu=0, \pm 1$ 
 to clarify the
role played by the non-trivial topological sector.   

To begin with, the formula for the quark condensate reads 

\begin{eqnarray}
\langle \bar{q} q \rangle & = & \langle \bar{u} u +\bar{d} d \rangle
=\frac{-1}{Z[J=m]} \left(  \frac{\delta Z}{
 \delta J_{uu}}|_{J=m}+(u\rightarrow d) \right)   \nonumber \\
 & = &  \frac{-1}{Z} 
 \int D[A]_{\nu =0} 
e^{-S_{YM}} {\rm Det}'[\rlap{/}{D} + m]  \times
 {\rm tr}[S(0,0)] 
\nonumber \\  &  & + \frac{-1}{Z}  \cdot 
\int D[A]_{\nu = \pm 1}
e^{-S^{YM}}{\rm Det}'[\rlap{/}{D} + m]  \times \\ \nonumber
 &  & \ \ \ \ \ \ \ \ \ \ \ \ \ \ \ \ \ \ \ \ \ \ \ 
 \ \ \ \ \ \ \ \ 
 \left[ \int d^4x \bar{\psi}_0(x) \psi_0(x) \cdot 
  \int d^4y \bar{\psi}_0(y)m_d \psi_0(y) +(u\rightarrow d) \right] ,
\end{eqnarray}
where $S(x,y) \equiv
 \langle x| ({\rlap{/}{D}+m})^{-1} |y \rangle  $.

 One finds that 
the first term in eq.(5) gives the Casher-Banks' formula\cite{CB},
\begin{eqnarray}
 \langle \bar{q}q \rangle
=- \pi \langle \rho(\lambda=0) \rangle ,
\end{eqnarray}
if one uses  the identity
\begin{eqnarray}
\label{den}
 {\rm tr}[S(0,0)] =\int d\lambda
 \rho(\lambda) \frac{2m}{\lambda^2+m^2} \stackrel{m\rightarrow 0}{
\longrightarrow}  \pi \rho(\lambda=0),
\end{eqnarray}
with $\rho(\lambda)$ being the density of fermion states with 
eigenvalue $\lambda$ ($ \rlap{/}{D} \Psi=i \lambda \Psi$) in the presence of
the gauge field $A_\mu$.

The second term of eq.(5) is from the $\nu= \pm 1$ sector and gives 
a correction of $O(m)$ to $\langle \bar{q} q \rangle $.  
Terms with higher $\nu$  have higher powers of $m$. 
For general $N_f$, the topological correction starts at $O(m^{N_f-1})$.
 Thus, in the chiral limit, all contributions from the
 $\nu \neq 0$  sector vanish 
and the Casher-Banks formula becomes exact. Namely 
 the Casher-Banks formula links chiral symmetry breaking 
 to the state density of fermions with nearly zero virtuality
  ($\lambda \sim 0$) coming only from the
 $\nu =0$ gauge configurations.  However,
 the fact that the exact zero modes in $\nu \neq 0$ sector
 cannot contribute to the formula in the chiral limit
 does not imply  that the topological configurations
 are irrelevant to the chiral symmetry breaking.  In the $\nu =0 $
sector, there exists many instanton + anti-instanton pairs which 
 could cause nearly zero modes: this can be explicitly seen in the
 chiral symmetry breaking in the phenomenological instanton-liquid model 
\cite{Shuryak2}.
  
 One should note here that
 the weighting function $e^{-S_{YM}} {\rm Det}'[\rlap{/}{D}+m]$ is
 positive semi definite.  Hence, eq.(5) implies that  
 the restoration of chiral symmetry  at high temperature ( i.e. vanishing
 of $\langle \bar{q}q \rangle $) is possible only when 
   $\rho(\lambda =0)$ 
  vanishes before averaging over the gauge configurations \cite{Tom1}.

Similar analysis with eq.(5) can be made for two point functions and higher.
Let us concentrate on the difference between $\sigma-\sigma$ and 
 $\delta-\delta$ two point functions.

\begin{eqnarray}
\langle \bar{q}(x)q(x)&,&\bar{q}(0)q(0) 
\rangle   
-  \langle \bar{q}(x)\tau^3 q(x),\bar{q}(0)\tau^3 q(0) 
\rangle \nonumber \\
 & = & \frac{1}{Z}  \int D[A]_{\nu =0}
e^{-S_{YM}}{\rm Det}'[\rlap{/}{D} + m]  
 \cdot {\rm tr}[S(x,x)] {\rm tr}[S(0,0)] \nonumber \\
 &  & + \frac{1}{Z}  
 \int D[A]_{\nu = \pm 1}
e^{-S_{YM}}{\rm Det}'[\rlap{/}{D} + m]\cdot 4  
 \cdot \bar{\psi}_0(x)\psi_0(x) \cdot \bar{\psi}_0(0) \psi_0(0)
+ O(m) \ \ .
\end{eqnarray}

Again, the first term comes from the $\nu=0$ sector and the second from the 
 $\nu= \pm1$. 
 In this case, the contribution from $\nu= \pm 1$ part remains even 
in the chiral limit.  At  high 
temperature when chiral symmetry gets restored, 
  $S(x,x)  \propto \rho(\lambda=0) \sim O(m)$.
  (A more detailed proof of this is given in ref.\cite{Tom1}.)
 From this, Cohen concludes that, when chiral symmetry gets restored, the 
 above difference vanishes and the $\sigma$ and $\delta$ become degenerate.
However, as can be clearly seen in eq.(8), the non-trivial topological 
configurations do not vanish and the difference remains.  
This result is quite general.  Consider taking the difference between 
two point functions such as $\langle J_1(x), J_1(0) \rangle-
\langle J_2(x),J_2(0) \rangle$.  Whenever we need a chiral rotation and
a $U(1)_A$ rotation to go from the spin zero current $J_1$  to $J_2$, 
non-trivial topological configurations contribute even in the chiral limit,
and the difference do 
not vanish even when chiral symmetry gets restored at high temperature.  

It should be noted, however, that for general $N_f$, effects of
 $\nu \neq 0$ sector 
  would be of $O(m^{N_f-2})$ and vanish in the chiral limit for $N_f \ge 3$.  
For   QCD in the real world,
  it is proportional to $m_s (\sim 150$ MeV) and the effect
 may not be negligible.

When the current $J_1$ and $J_2$ have spin-1 or 2, nontrivial 
topological contributions vanish due to the chirality of the fermion 
zero mode irrespective of $N_f$ \cite{GI}.  

For general  n-point functions (n$\ge3$) or their differences, 
there again is going to
be a non-trivial contribution from $\nu \neq 0$ sector.
 Generalization to higher $N_f$ in this case is also easy.
   One immediately finds  
that the topologically non-trivial configurations start to contribute from
n-point functions where n=$N_f$.

Let us now  summarize our main conclusions:

\begin{enumerate}

\item  $\nu \neq 0$ sectors  do not contribute 
 to two point functions for $N_f\ge$ 3 in the $SU(N_f)$ symmetric chiral 
limit.  In this case, the physical spectrum will effectively be multiplets of
 $U(N_f) \times U(N_f)$, as was claimed in \cite{Tom1}.

\item If $N_f=2$, $\nu \neq 0$ sectors do contribute
differently to the two point functions of 
  $\pi,\delta,\sigma,\eta $  even in the chiral limit and 
even when chiral symmetry is restored at high temperature.  
In this case,  argument in \cite{Tom1} does not apply.

\item In the real world,  $SU(3)$ is not exact and $m_s \sim 150$MeV.  
In this case,  $\nu \neq 0$ sectors do contribute with possible
 non-negligible effect even at high temperature.

\end{enumerate}

Quantitative estimates of the
magnitude of $\nu \neq 0$ sectors in case 2 and 3 at high temperature
 depend on the density of topological configurations at high temperature.
 Although such density  could be highly suppressed 
  and $U_A(1)$ symmetry 
 could be {\em effectively} restored at high temperature,
 our arguments in this letter show that there is always a finite amount of
 $U_A(1)$ symmetry breaking in the particle spectrum even when
  chiral symmetry restoration occurs at high temperature.

\acknowledgements

We would like to thank T. D. Cohen for explaining to us his work before
submission to the hep-ph bulletin board.  SHL was supported in part by the 
JSPS short term visitor fellowship, by the Basic Science Research
Institute program of the Korean Ministry of Education through grant no.
BSRI-95-2425 and by KOSEF through the CTP at Seoul National University. 
TH was supported 
 in part by the Grants-in-Aid of the Japanese Ministry of 
Education, Science and Culture (No. 06102004).


\begin{thebibliography}{99}
\bibitem{note} We have implicitly assumed 
  $n_+=1$ in this formula. When  $n_+ > 1$ for $\nu =1$, 
$Z[J]_{\nu=1}$ has extra powers of $J$ and hence it is irrelevant
 for our argument below.
\end{thebibliography}
\end{document}